\documentclass{aa}

\usepackage{graphics}

\def\simless{\mathbin{\lower 3pt\hbox
{$\rlap{\raise 5pt\hbox{$\char'074$}}\mathchar"7218$}}}   
\def\simmore{\mathbin{\lower 3pt\hbox
{$\rlap{\raise 5pt\hbox{$\char'076$}}\mathchar"7218$}}}   

\begin{document}


\title{Orbital Comptonization in accretion disks around black holes}

\author{Pablo Reig\inst{1,2}
\and Nikolaos D. Kylafis \inst{1,2}
\and Henk C. Spruit \inst{3}
}

\institute{University of Crete, Physics Department, P.O. Box 2208, 710 03,
Heraklion, Crete, Greece
\and Foundation for Research and Technology-Hellas, 711 10, Heraklion,
Crete, Greece
\and Max Planck Institute for Astrophysics, Box 1317, D-85741 Garching,
Germany
}

\offprints{pablo@physics.uoc.gr}

\date{Received / Accepted}

\abstract{
We have performed Monte Carlo  simulations of Compton upscattering  of
low-energy photons in an accretion disk around a Schwarzschild black
hole.  The photons gain energy from the rotational motion of the electrons in
the disk. The  upscattering  occurs  near the black  hole horizon, where
the  flow velocity of the  electrons approaches the speed of light.  We
show that this type of bulk-flow Comptonization can  produce  power-law
X-ray spectra  similar to the ones observed in black-hole  X-ray
transients in the high/soft state, i.e., a soft bump dominating the
spectrum below $\sim$ 10 keV and a power-law tail with photon index in the
range 2--3.  In order to reproduce the observed hard to soft flux ratio
the disk has to have vertical
optical depth above $\sim 3$ at the last stable orbit.  We conclude
that the power-law component of the high/soft state of black-hole
transients may be due to an intrinsically cool disk extending all the
way to the hole, without a separate hot plasma component.
}

\maketitle

\section{Introduction} \label{intro}

The X-ray continuum spectrum of black-hole binaries is characterized by an
ultrasoft component and a power-law hard tail. The ultrasoft component is
interpreted as coming from an optically thick accretion disk.  The
temperature of the disk is high ($\sim 1$ keV)
near the black hole and decreases
outwards.  Thus, the spectrum emitted by the disk is a multi-temperature
black-body spectrum. The power-law tail is thought to originate from
Comptonization of soft photons scattering off very energetic electrons.
The amplitude of this component has been seen to correlate with the $1 -
10$ keV flux. During the low (also called hard) state, the ultrasoft
component is very weak or absent, whereas in the high (soft) state it
dominates the spectrum, especially at energies $\simless 10$ keV.

Two types of Comptonization processes have been put forward to explain the
power-law component: thermal and bulk-flow Comptonization. Thermal
Comptonization requires the presence of a hot and rarefied cloud of
electrons or {\em corona} (see, e.g.,  the review by Poutanen 1998 and
references therein), where the soft photons get upscattered due to
collisions with Maxwellian distributed 50-100 keV electrons. In the {\em
bulk-flow} Comptonization scenario (see, e.g., the review by Kylafis \&
Reig 1999 and Papathanassiou \& Psaltis 2001),  photons gain energy from
collisions with radially infalling electrons. Titarchuk et al. (1997)
assessed the relative importance  of these two processes and concluded
that the bulk  motion is more efficient  in  upscattering   photons than
thermal Comptonization in spherical accretion, provided  that  the
electron temperature  in the radial flow is less than a few keV.

Although these  Comptonization  models have been succesful in describing
the spectra of black-hole candidates, there are still a number of
unresolved issues. A radial inflow of  plasma  has been considered for
mathematical simplicity, but how this radial inflow is formed is unclear.
Since the disk is likely to have a magnetic field generated by the
shear flow, it may well have a corona heated by magnetic dissipation
(Galeev et al. 1979). Comptonization in this hot corona is one of the
currently popular models for the hard spectral components in X-ray
binaries, the other being two-temperature flows or `ADAFs'. There is
little independent evidence for either of these possibilities, however,
apart from the interpretation of the X-ray spectrum they offer.

On the other hand, general agreement exists on the presence of the
accretion disk itself around a black hole in a  mass transfering  binary
system.  In this work we investigate the Comptonization process in a cold
accretion disk around a black hole by Monte Carlo simulations. Soft
photons produced inside or outside the disk get upscattered by collisions
with fast-moving electrons, which follow circular trajectories outside the
last stable orbit and ballistic trajectories inside it.  We refer to this
process by the name `orbital Comptonization'. It has been considered
before by Hanawa (1990) who applied it to the boundary layer between a
disk and a neutron star surface.

\begin{figure}
\mbox{}
\vspace{6.0cm}
\includegraphics{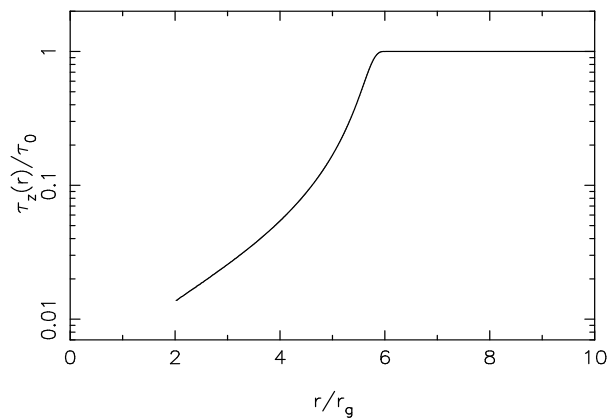}
\caption[]{Variation of the vertical optical depth with radius in the inner
disk
(normalized to the optical depth $\tau_0$ in the outer parts of the disk),
for the
model used in the scattering calculations.}
\label{tau}
\end{figure}

\subsection{Thermal vs.\ bulk-flow Comptonization}

In a geometrically thin disk, the scattering photons escape easily before
they have
a chance to pick up energy from the orbital motion. Bulk-flow Comptonization
is
therefore likely to work better in thicker disks. Such disks are also hotter
however, so that thermal Comptonization will also take place.

To compare the relative contributions of the two processes, note that the
scattering photons drift from their point of origin by random walk. Before
they escape through the disk surface, they therefore sample the orbital
velocity field over  a radial distance which is of the same order as the
disk thickness. The typical  velocity difference encountered by a
scattering photon is thus of the order  $\bar{v} = \Omega H\sim c_{\rm s}$,
where $c_{\rm s}$ is the sound speed in the disk and $\Omega$ is the
angular velocity.  In the absence of radiation pressure, $c_{\rm s}\sim
(k_{\rm B}T/m_{\rm p})^{1/2}$. In thermal equilibrium, the thermal
velocity of the scattering electrons is $c_{\rm s}(m_{\rm p}/m_{\rm
e})^{1/2} \sim(m_{\rm p}/m_{\rm e})^{1/2} \, \bar{v} \gg \bar{v}$. It would
thus seem that thermal Comptonization wins over bulk-flow Comptonization
by the orbital motion. For cool Shakura-Sunyaev disks this would indeed be
the case. However, the temperatures in such disks, of the order 1 keV, are
not sufficient for significant Comptonization anyway.

More interesting are the geometrically thick accretion flows that are
possible in the inner regions of disks. Here the conditions are different
from those in standard thin disks. In the optically thin ADAF type flows,
thermal equilibrium between protons and electrons does not hold, and the
scattering electrons are much colder than expected from the thickness of
the disk. The electron temperatures are near  100 keV so that the electron
velocities are around $0.5c$. This is of the  same order as the range of
orbital velocities sampled by a photon escaping through a disk of
thickness $H/r \sim1$ near the last stable orbit. In the optically thin
type of ADAF, orbital Comptonization can therefore be as important as
thermal Comptonization, though it may not dominate.

Another type of radiatively inefficient accretion flow into a black hole
takes place when the optical depth of the flow is large, and the disk is
puffed up to aspect ratios $H/r \sim 1$ by radiation pressure. The
temperature in such radiatively supported disks is relatively low, so that
thermal Comptonization is not very effective, but the escaping photons
will have sampled a large range of orbital velocities. The conditions in
optically  thick radiation supported flows such as may be present at high
accretion rates onto black holes are thus especially relevant for orbital
Comptonization.

\section{The model}

\subsection{Densities and velocity field}

In an accretion disk around a non-rotating black hole of mass $M$ matter
follows nearly circular orbits only up to the last stable orbit at a
distance of $r_{\rm last}=6 r_{\rm g}$, where $r_{\rm g}=GM/c^2$ is  the
gravitational radius.  At the black hole horizon $r_{\rm h}=2 r_{\rm g}$,
the speed of the particles is $c$.

The details of bulk Comptonization depend on the velocity field of the
accretion flow, as well as its optical depth and geometrical thickness as
a function of the distance from the hole. For the high accretion rates we
have in mind, the flow has a finite thinkness of the order $r_{\rm g}$,
depending on the details of disk viscosity and radiative transfer. Instead
of a more detailed quantitative model, we approximate the velocity field
by purely circular motion outside the last stable orbit, and a ballistic
spiral-in inside $r=r_{\rm last}$, as follows:

For $r \ge r_{\rm last}$, matter in the disk is taken to be in circular
motion. In Schwarzschild coordinates, the orbital frequency as seen by a
distant inertial observer is then $\Omega=(GM/r^3)^{1/2}$ [Misner, Thorne,
\& Wheeler 1973, (hereafter MTW) p671], the same as in Newtonian gravity.
At $r=r_{\rm last}$ and in geometric units, the energy of  the particles
per unit  mass is $E=\sqrt{8/9}$ and their angular momentum per unit mass
$L$ divided by the black-hole mass $M$ is $L/M = \sqrt{12}$ (MTW p662).

To find the velocity of the gas as a function of $r$ as it falls
ballistically into the black hole,  we use the expressions (Shapiro \&
Teukolsky 1983 pp341-342)

\begin{eqnarray}
v_{\phi}(r) &=& \frac{L}{rE}\sqrt{1-\frac{2M}{r}} \\
v_{r}(r) &=& \frac{1}{E}\sqrt{E^2-\left(1-\frac{2M}{r}\right)
\left(1+\frac{L^2}{r^2}\right)}
\label{vr}
\end{eqnarray}

\noindent We assume that just inside $r_{\rm last}$, $E=\sqrt{8/9}$ and
$L/M = \sqrt{12} - \epsilon$, where $\epsilon \ll \sqrt{12}$ is a measure
of the  radial drift speed in the accretion disk just outside the last
stable orbit.  The exact value of $\epsilon$ has no effect on the
calculations.

To compute the density and the vertical optical thickness of the flow, we
assume a constant thickness $H=2 r_{\rm g}$ (as seen by a distant
observer). From the continuity equation and the assumed velocity field,
the vertical optical depth can then be computed from the radial velocity
(Eq.~\ref{vr}). Outside $r_{\rm last}$, where we have assumed circular
orbits,
we take a constant density. The resulting vertical optical depth of the
disk as a function of distance, $\tau_z(r)$, is shown in Fig.~\ref{tau}.

Bulk-flow Comptonization that produces hard X-rays is effective only near
the black-hole horizon, where  the speed of the electrons is high.  Thus,
only those photons that find themselves  in the inner part of the disk,
and manage to escape eventually, will have their energy significantly
increased.  It is these photons that produce the high-energy power-law
spectra.

\begin{figure}
\mbox{}
\vspace{12.0cm}
\includegraphics{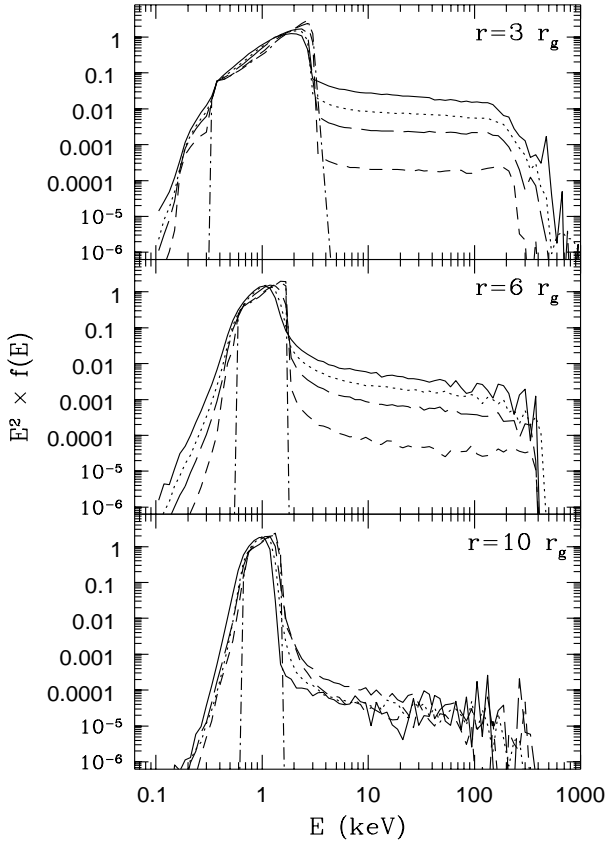}
\caption[]{Emergent spectral energy distributions as functions of position
$r/r_{\rm g}$ of the source of soft photons (monochromatic at 1 keV, in
the midplane of the disk), and $\tau_0$, a measure of the vertical optical
thickness (see text and Figure 1). $\tau_0 = 0$ (dot-dashed), $0.1$
(short-dashed), $1$ (long-dashed), $3$ (dotted), and $10$ (solid). The
three panels are for $r=3 r_{\rm g}$ (top), $r=6 r_{\rm g}$ (middle) and
$r=10 r_{\rm g}$ (bottom).}
\label{mono}
\end{figure}

\begin{figure}
\mbox{}
\vspace{12.0cm}
\includegraphics{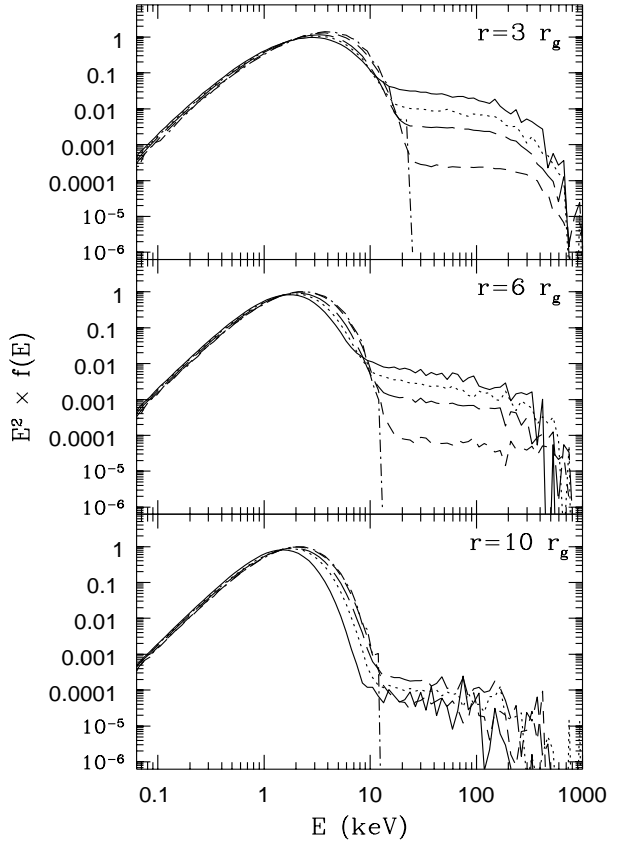}
\caption[]{Same as in Fig.~\ref{mono}, but with a blackbody source
(dot-dashed line)
of soft photons of temperature $kT=0.5$ keV instead of a monochromatic
source.}
\label{bbody}
\end{figure}

\begin{figure}
\mbox{}
\vspace{6.0cm}
\includegraphics{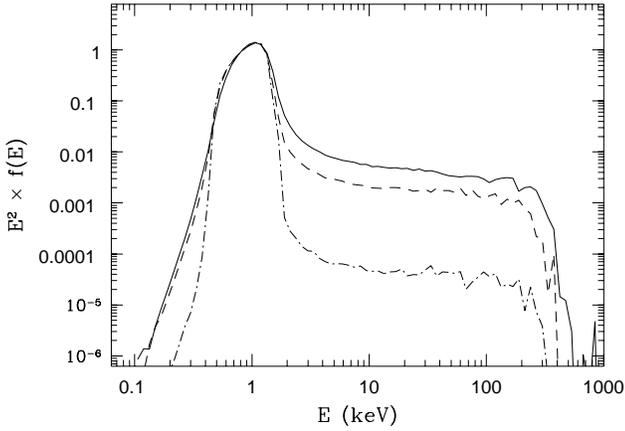}
\caption[]{Dependence of the spectrum on disk thickness,
$H~=~0.3~r_{\rm g}$ (dot-dashed line), 1 $r_{\rm g}$
(dashed line) and 5 $r_{\rm g}$ (solid line). The input source of photons
is monochromatic at 1 keV and is located at 6 $r_{\rm
g}$. The  optical depth was taken to be 3}
\label{thick}
\end{figure}

\section{Radiative transfer}

Photons are generated throughout the flow, diffuse over some distance and
leave it at some other point. The radiation leaving the disk at any point
of its surface is thus an integral over the contributions from all source
positions. In order to highlight the importance of different source
positions (in particular, of their distance to the hole), we show here
results for fixed source positions. The net spectrum can be obtained from
this by integration, if needed. The emission of photons at the source is
assumed to be isotropic in a comoving frame, as it would be for photons
generated by bremsstrahlung, for example.

As a photon  travels in the disk, it  experiences   Compton scatterings
with the orbiting electrons.  If the optical depth is small ($\simless
1$), the majority of the input photons  escape  unscattered.  Those that
are directed towards the black hole get absorbed. Those that scatter a few
times have on average their energy significantly increased.  If the
optical depth is moderate, then those photons that are not absorbed by the
black hole random-walk  through the medium prior to escape, and gain
energy from the bulk motion of the electrons.

We make some simplifications in dealing with the general relativistic
aspects of the problem. The effect of curvature on the path of the photon
between scatterings is neglected, that is, we approximate it as a
straight-line as seen by a distant observer. Since the distances traveled
between scatterings are small, this is not a bad approximation.  For a
quantification of the small error  made with the use of straight-line
photon trajectories the reader is referred to Papathanassiou \& Psaltis
(2001).  The effect of curvature on the geometrical distances is also
neglected since the actual physical distance between scatterings is not
important for the scattering process, only the physical conditions at the
scattering centers. Finally, we leave out the self-illumination of the
disk surface by other parts of the disk (through gravitational light
bending, for example).

We do take into account the important effect of the changing gravitational
redshift along the photon trajectory. The energy $E(r_1)$ of a photon, as
measured by a locally inertial observer at radius $r_1$, is related to the
energy of the same photon $E(r_2)$ at radius $r_2$, as measured by a
locally inertial observer through the difference in gravitational redshift
(MTW p659)

\begin{equation}
E(r_1) \sqrt{1 - 2r_{\rm g}/r_1}
= E(r_2) \sqrt{1 - 2r_{\rm g}/r_2}
\end{equation}

\subsection{Monte Carlo implementation}
The basic  principle  of the  Monte  Carlo  technique, as it is applied
to the Comptonization  process (Cashwell \& Everett 1959; Pozdnyakov,
Sobol, \& Sunyaev 1983),  consists of following  the life histories of a
large number of photons  from the moment each photon is emitted  from a
given  source  until it leaves the  scattering  medium.

The photon is characterized by 4 parameters (position,  direction, energy
and  weight) which are updated at each scattering, and determine the
spectrum of the radiation  emerging from the  scattering medium.  The
procedure can be divided into the following steps:

\begin{enumerate}

\item The disk is divided into concentric cylindrical zones. Each
zone is characterized by a mean velocity and a mean density. The number of
zones is chosen so as to have a fine zoning of electron velocities in the
inner parts of the disk.

\item We calculate the optical depth to electron scattering  $\tau
(\vec{r}, \vec{n}, E)$ from the scattering position $\vec{r}$
to the boundary of the disk
along the direction of the photon ${\bf \vec{n}}$. Then, a portion $W
e^{-\tau}$ of the photon's weight escapes and it is recorded, while the
remaining weight  $W^{\prime} = W (1- e^{-\tau})$ is scattered between
$\vec{r}$ and the boundary.

\item We determine the new scattering position, the energy of the
photon after scattering and its new direction.

\end{enumerate}

The  process  continues  with the next  scattering  until the photon's
weight falls below a certain  value   ($10^{-8}$ in our runs),  and it is
repeated for $N$ photons, where $N$ is a large number (typically $10^7$).

\begin{figure}
\mbox{}
\vspace{6.0cm}
\includegraphics{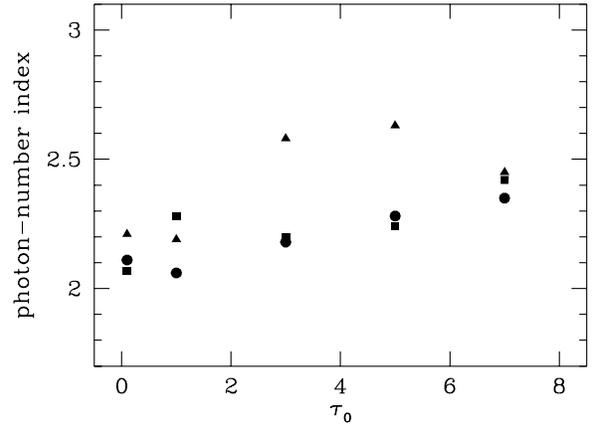}
\caption[]{Photon-number
spectral index in the range 20--100 keV
as a function of vertical optical depth $\tau_0=\tau_z(r_{\rm last})$
for the spectra shown in
Fig.~\ref{mono} top (circles),  middle (squares) and bottom (triangles).}
\label{index}
\end{figure}
\begin{figure}
\mbox{}
\vspace{6.0cm}
\includegraphics{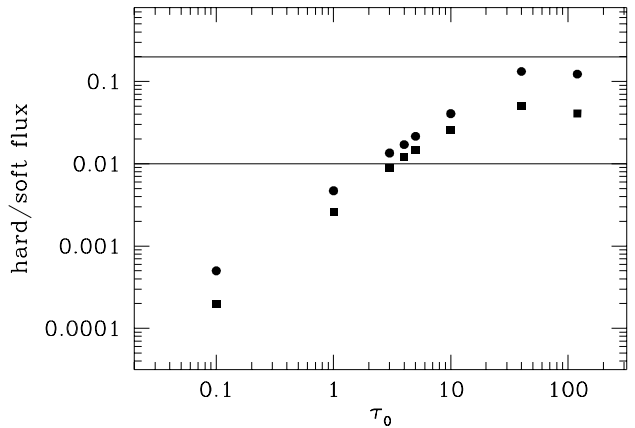}
\caption[]{Hard to soft flux ratio as a function of the vertical optical
depth
$\tau_0=\tau_z(r_{\rm last})$ and source
position: $r=3 r_{\rm g}$ (circles), $r=6 r_{\rm g}$ (squares). The hard and
soft bands are 20--200 keV and 1--5 keV, respectively.
The solid lines enclose the region compatible with observations.}
\label{ratio}
\end{figure}

\section{Results and discussion}

We consider a point source of soft photons in the
midplane of the accretion disk.  In the local rest frame of the flow, the
source emits monochromatic photons of  energy $E_0 = 1$ keV
isotropically.  In the frame of a local inertial observer the distribution
of  emitted photons is not isotropic but forward peaked (in the direction of
the flow) and the initial energy of the photons is Doppler shifted
and  depends on their initial direction.

The resulting spectral energy distributions $E^2f(E)$, where $f(E)$ is the
photon-number spectrum, are shown in Figs.~\ref{mono} -- \ref{thick}. 
Figure~\ref{mono} shows results  for three distances of the soft photon
source from the hole, $r/r_{\rm g}$ = 3,  6 and 10, and for five values of
the vertical optical depth at the last stable orbit. All five spectra in
each panel are normalized so that the integral of $f(E)$ from zero to
infinity is equal to 1.

The emergent spectra in Fig.~\ref{mono} consist of a peak, whose width
increases as the source of photons approaches the black hole,   and a
high-energy tail that is approximately a power law with a cut off at 100--
300 keV. The soft peak is formed by those photons that escape unscattered
or have had a small energy change before escape.  The tail is prominent
and extends to high energies even if the optical depth is low. It is
formed by those few photons that picked up a lot of energy from the
electrons, mainly through head-on collisions.

Since model fits to the observed X-ray spectra in black-hole candidates
give  blackbody temperatures for the accretion disk in the range $\sim$
0.3 -- 1 keV (Tanaka \& Lewin 1995), we have also considered a source of
soft photons in the disk with blackbody emission of temperature $kT = 0.5$
keV.   The pointlike source is again placed at the midplane of the disk. The
emergent spectra are shown in Fig.~\ref{bbody}.  The panels and lines
have the same meaning as in Fig.~\ref{mono}.

The amplitude of the high-energy tail and the cutoff photon energy
decrease with distance of the photon source from the hole. This is
expected since the orbital velocities encountered by the photons decrease
with distance.   The photon-number spectral index is, however, less
dependent on the source position. Within statistical errors, it is also
the same for a monochromatic and a black body input spectrum.

Since the scattering photons sample a velocity range that depends on the 
disk thickness (cf. the discussion in section 1.1), the spectra depend not
only on the optical depth, but also on the actual geometric thickness of the
disk. This is illustrated in Fig.~\ref{thick}. It shows that the high energy
tail is strongly suppressed when the aspect ratio $H/r$ drops substantially
below unity. 

These spectra are reminiscent of the observed spectra in black-hole
candidates in the high/soft state. The resemblance is strengthened when
the photon-number spectral index  and the ratio of the hard to the soft
flux are compared with observations.

The photon-number spectral index in the tail varies in the range 2--3
and is, within statistical errors, independent of the position of the
source of soft photons. It depends on the vertical optical depth
$\tau_0$ at $r=r_{\rm last}$;   the thicker the medium the softer the
spectrum. Figure~\ref{index} shows the dependence of the power-law
photon-number  spectral index on the vertical optical depth $\tau_0$.
The index was computed from the spectra of Fig.~\ref{mono}, in the energy
range between 20 and 100 keV. For low and intermediate optical depths, the
photon-number index is $\sim 2-2.5$, which is similar to those observed in
the high/soft state of black-hole candidates (Tanaka \& Shibazaki 1996).

In addition to the slope of the high-energy tail, we can also compare its
strength with the observations. We do this in terms of a hard-to soft flux
ratio.   Observations of soft state spectra (e.g. Miyamoto et al. 1993;
Cui et al. 1997),  generally give values of 0.01--0.2 for the ratio of
power law tail to black body fluxes. In Fig.~\ref{ratio} we have plotted
the 20--200 keV to 1--5 keV flux ratio of our models as a function of the
vertical optical depth $\tau_0$.

In a realistic accretion disk, it is expected that the majority of the
photons in the ultrasoft component come from the region of the disk near
$6r_{\rm g}$.  Close to the black-hole horizon the vertical optical depth
of the disk is relatively small and at $r \gg 6r_{\rm g}$ the temperature
and emissivity of  the disk are small.  Thus, the squares in
Fig.~\ref{ratio} are closer to reality than the circles.  In other words,
the vertical optical depth of the disk at the last stable orbit must be
greater than $\sim 3$ in order for our proposed mechanism to produce
spectra similar to the ones observed.

The energy of the photons escaping to infinity is the result of the
competition of two processes: the gain of energy due to Compton
scattering, which increases towards the horizon, and the loss of energy
due to  gravitational redshift, which also increases.  It can be shown
that for a thin disk the escaping photon energy cannot exceed $m_{\rm
e}c^2$.  Some of the spectra in Fig.~\ref{mono} and \ref{bbody} show a
cutoff at energies a bit above this limit. This can be traced back to a
simplification made in our assumed velocity field. The rotational velocity
was taken the same at all $z$ (for the same polar coordinate $\rho$). On
the other hand, in the gravitational redshift the radial coordinate enters
spherically, i.e. $r = \sqrt{\rho^2 + z^2}$, which is larger than $\rho$.
In reality, the rotation speed will decrease in step with the redshift.

In conclusion, we have discussed for the first time the Comptonization by
orbital motion in an accretion disk around a Schwarzschild black hole (as
opposed to thermal Comptonization in a hot corona or bulk-motion
Comptonization by radially-falling electrons) and have shown that
high-energy power-law X-ray spectra of the type seen in the soft/high
states of black hole accreters.

\begin{acknowledgements}
The authors acknowledge partial support from the European Union Training
and Mobility of Researchers Network Grant ERBFMRX/CT98/0195.
\end{acknowledgements}
\end{document}